\documentclass[aps,prd,twocolumn,showpacs,superscriptaddress,groupedaddress,10pt]{revtex4-2}
\usepackage{graphicx}  % needed for figures
\usepackage{dcolumn}   % needed for some tables
\usepackage{bm}        % for math
\usepackage{amssymb}   % for math

\usepackage{amsfonts}
\usepackage{amsmath}
\usepackage{bm}
\usepackage{array}
\usepackage{amssymb}
\usepackage{array}
\usepackage{amsthm}
\usepackage{amssymb}
\usepackage{braket}
\usepackage{color}

\newcommand{\be}{\begin{eqnarray}}
\newcommand{\ee}{\end{eqnarray}}
\renewcommand{\d}{\mbox{${\rm d}$}} %d differenziale non corsivo in math mode

\newcommand{\K}{\mathcal{K}}
\newcommand{\T}{\mathcal{T}}

\begin{document}

\title{Revisiting induced gravity in scalar-tensor thermodynamics}
\author{Andrea Giusti}
%\email{andrea.giusti9@unibo.it}
\affiliation{DIFA \& ${\cal AM}^2$, University of Bologna, 40126 Bologna, Italy}
\affiliation{I.N.F.N., Sezione di Bologna, I.S.~FLAG}

%\date{\today}

\begin{abstract}
\noindent Induced gravity, defined as a globally scale-invariant ``first-generation'' scalar-tensor theory, is investigated within the framework of the thermodynamics of modified gravity theories. 
The ``temperature of gravity'' and its evolution equation are derived for this model, and the resulting expressions are used to analyse General-Relativity equilibrium states and to investigate the possible existence of an attractor mechanism toward Einstein's theory with a cosmological constant.
\end{abstract}

\maketitle
\section{Introduction}
\label{sec:intro}
\noindent The generation of Einstein's gravity through processes of spontaneous symmetry breaking has been a topic of general interest since the early 1980s \cite{Sakharov:1967nyk, Zee:1978wi, Cooper:1981byv, Zee:1980sj, Turchetti:1981zb}. A very well-known approach along this line relies on the spontaneous breaking of scale invariance within the framework of scalar-tensor gravity
\cite{Zee:1978wi, Cooper:1981byv, Zee:1980sj, Turchetti:1981zb}. Concretely, the most general ``first-generation'' globally scale-invariant scalar-tensor theory, originally proposed by F.~Cooper and G.~Venturi in \cite{Cooper:1981byv} and commonly known as {\em Induced Gravity}, reads \footnote{In this work we adopt the notation of Ref. \cite{Wald:1984rg}, in which the metric signature is $(-+++)$. Furthermore, units are used in which the speed of light and $8 \pi G$ (where $G$ denotes Newton's constant) are unity.}
\be
\label{eq:CV}
\mathcal{S}[g,\phi] &\!\!=\!\!& \int \d ^4 x \sqrt{-g} 
\left[ \frac{\xi}{2} \, \phi^2 \, R - \frac{1}{2} \, \nabla_a \phi \nabla^a \phi - \frac{\lambda}{4} \, \phi^4 \right] \nonumber \\
&& + \mathcal{S}^\mathrm{(m)} \, ,
\ee 
where $\lambda, \xi > 0$ are dimensionless constants, $\phi$ is a scalar field, $g$ is the determinant of the spacetime metric $g_{ab}$ with Ricci scalar $R$, $\nabla$ is the  Levi-Civita connection and $\mathcal{S}^\mathrm{(m)}$ denotes the action for matter fields (assumed to be independent of $\phi$).

The field equations for the action in \eqref{eq:CV} read
\be
&& \Box \phi + \xi R \phi - \lambda \phi^3 = 0, \label{eq:CVeq1} \\[1mm] 
&& \xi \phi^2 \, G_{ab} = T^{(\mathrm{m})}_{ab} + \nabla_a \phi \nabla_b \phi - \frac{1}{2} g_{ab} \nabla_c\phi \nabla^c\phi \nonumber \\
&& \qquad \qquad \quad- \frac{\lambda}{4} \, \phi^4 \, g_{ab}  + \xi \big( \nabla_a \nabla_b \phi^2 - g_{ab} \, \Box \phi^2 \big) \, , \label{eq:CVeq2} \qquad
\ee
with $\Box := g^{ab} \nabla_a \nabla_b$ the Laplace-Beltrami operator, $G_{ab} := R_{ab}-({R}/{2}) \, g_{ab}$ the Eintein tensor, $R_{ab}$ the Ricci tensor associated with $g_{ab}$, and $T_{ab}^\mathrm{(m)} $ denoting the matter stress-energy tensor.

It is worth pointing out (see also Ref. \cite{Finelli:2007wb}) that this model can be mapped onto the following realization of Brans-Dicke (BD) theory
\be 
\label{eq:BDA}
\mathcal{S}[g,\psi] &\!\!=\!\!& \frac{1}{2}\int \d ^4 x \sqrt{-g} 
\left[ \psi \, R - \frac{1}{4\xi} \, \frac{\nabla_a \psi \nabla^a \psi}{\psi} - \frac{\lambda}{2 \xi^2} \, \psi^2 \right] \nonumber \\ 
&& + \mathcal{S}^\mathrm{(m)}\, ,
\ee
{\em via} the field redefinition
\begin{equation}
\label{eq:field-red}
\psi = \xi \phi^2 \, ,
\end{equation}
thus resulting in a BD model with constant coupling $\omega = 1/(4 \xi)$ and potential $V(\psi) = \lambda \psi^2/(2 \xi^2)$. Notably, the condition $\xi > 0$ allows the model to avoid pathological behaviours such as phantom phenomenology $(2\omega + 3 < 0)$ and a non-dynamical BD scalar $(2\omega + 3 = 0)$.

Taking advantage of this mapping onto BD theory, we can briefly revisit the discussion in \cite{Cooper:1981byv} concerning the spontaneous breaking of scale invariance in a cosmological setting. 

Consider a flat Friedmann-Lema\^{i}tre-Robertson-Walker (FLRW) spacetime, with line element
$$
\d s^2 = - \d t^2 + a^2(t) \left( \d r^2 + r^2 \d \Omega^2 \right)
$$
where $t$ is cosmic time, $a(t)$ is the scale factor, $r$ denotes the comoving radial coordinate, and $\d \Omega^2$ is the  line element on the unit $2$-sphere. Restricting our discussion to the {\em vacuum} case we find that the dynamics of the scale factor and of the scalar field $\psi = \psi (t)$ are determined by the scalar field equation and the Friedmann equation, {\em i.e.} (see \cite{Faraoni:2004pi})
\be 
&& \ddot{\psi} + 3H \dot{\psi} = 0 \, , \label{eq:psi}\\
&& H^2 = \frac{\omega}{6} \frac{\dot{\psi}^2}{\psi^2} - H \frac{\dot{\psi}}{\psi} + \frac{V_0 \, \psi}{6} \label{eq:friedmann} \, ,
\ee
with $\omega = 1/(4 \xi)$, $V_0 = \lambda/(2 \, \xi^2)$, $H := \dot{a}/a$, and the dot denoting the derivative with respect to $t$. 

It is possible to show (see \cite{Cooper:1981byv,Faraoni:2004pi}) that this system has no power-law solutions, except for the trivial one with
\be 
\psi = \psi_0 \quad \mbox{and} \quad H^2 = \frac{\lambda \, \psi_0}{12 \, \xi^2} \, ,
\ee 
for any $\psi_0 > 0$ constant. This solution of {\em vacuum} BD gravity leads to the well-known {\em de Sitter solution} discussed in \cite{Cooper:1981byv,Turchetti:1981zb}, {\em i.e.}
\be 
\label{eq:dSVacuum}
\phi = \phi_0 \quad \mbox{and} \quad H = H_0 = \sqrt{\frac{\lambda \, \phi_0^2}{12 \, \xi}} \, ,
\ee
in the induced gravity theory, by taking advantage of the field redefinition in \eqref{eq:field-red}. Notably, although the field equations of the theory are scale invariant, these {\em vacuum} solutions are not. Indeed, by selecting a value $\phi_0 \neq 0$, the scalar field acquires a non-vanishing {\em vacuum expectation value}  $\braket{\phi} = \phi_0 \neq 0$, and hence the {\em vacuum} is not scale invariant. In other words a spontaneous breaking of scale invariance occurs, leading to the generation of the Planck scale $M_{\rm P}^2 = \xi \phi^2 _0$. 

Induced gravity coupled with dust and radiation has been shown to have 
General Relativity (GR) plus cosmological constant as a stable attractor \cite{Finelli:2007wb} for a range of initial conditions. This scenario reminds of the general attractor-to-GR mechanism in scalar-tensor gravity proposed by T.~Damour and
K.~Nordvedt (see \cite{Damour:1992kf,Damour:1993id}).

In this work we revisit the induced gravity paradigm by taking advantage of the recently developed framework of the thermodynamics of 
scalar-tensor gravity (see \cite{Faraoni:2018qdr,Faraoni:2021lfc,Faraoni:2021jri,Giusti:2021sku,Giardino:2023ygc}).

\section{Thermodynamics of Scalar-Tensor Gravity: an overview}
\label{sec:thermo}
In this section we recall some essential notions of the so-called thermodynamics 
of scalar-tensor theories. This discussion will be based on \cite{Faraoni:2018qdr,Faraoni:2021lfc,Faraoni:2021jri,Giusti:2021sku,Giardino:2023ygc} (specifically, \cite{Giardino:2023ygc} offers an introductory review of the general approach), to which we refer the interested reader for further details on the formalism. We will also further specialise the discussion to cosmological spacetimes thus recalling some ideas from \cite{Giardino:2022sdv, Miranda:2024dhw} and from \cite{Faraoni:2025alq}, the latter concerning the attractor-to-GR mechanism in scalar-tensor gravity. 

Consider the action of {\em viable Horndeski gravity} coupled to matter
\begin{equation}\label{eq:actiongen}
\mathcal{S}\left[g_{ab},\psi, \zeta\right]=\mathcal{S}^{\rm(g)}\left[g_{ab},\psi\right]+\mathcal{S}^{\rm(m)}\left[g_{ab},\zeta\right]\,,
\end{equation}
with
\begin{equation}\label{eq:actionH}
\mathcal{S}^{\rm (g)}=\frac{1}{2}\int \d ^4 x \sqrt{-g} \, \Big[ G_{4} \, R +G_{2}-G_{3}\Box\psi\Big] \, ,
\end{equation}
where $G_i$ are arbitrary functions of $\psi$ and/or $X := -(1/2) \nabla^a \psi \nabla_a \psi$ and with the requirement that $G_{4X} = 0$, adopting the convention $f_{\psi}  \equiv \partial f/\partial \psi$ and $f_{X} \equiv \partial f/\partial X$.

The field equations for viable Horndeski gravity can be rewritten as (see \cite{Giusti:2021sku, Miranda:2024dhw, Gallerani:2024gdy})
\begin{align}
& G_{ab} = G_{\rm eff} (\psi) \, T^{\rm (m)} _{ab} + T^{\rm (eff)} _{ab} \, , \label{fe1} \\
& \frac{\delta S^{(g)}}{\delta \psi} = 0 \, , \label{fe2} \\
& \frac{\delta S^{(m)}}{\delta \zeta} = 0 \, , \label{fe3}
\end{align}
{\em i.e.} they split into an effective Einstein equation, an equation for the scalar field $\psi$, and the field equations for the matter fields $\zeta$. Note that the effective Einstein equation has two contributions on the right-hand-side. The first contribution is due to the stress-energy tensor of matter which couples to (Einstein) gravity {\em via} an effective gravitational coupling, that for the case of viable Horndeski gravity is simply $G_{\rm eff} (\psi) = 1/ G_4 (\psi)$. The second term $T^{\rm (eff)} _{ab}$ denotes an effective stress-energy tensor accounting for the rest of the contributions of $\psi$ to the effective Einstein equation.

If the $4$-gradient of $\psi$ is timelike, we can construct 4-velocity field
$$
u^a := \epsilon \, \frac{\nabla^c\psi}{\sqrt{2X}} \, , \quad 2X = -\nabla^e \psi \nabla_e\psi \, ,
$$
where $\epsilon = \pm 1$ has to be chosen so that $u^a$ is future-oriented by definition. The stress-energy tensor $T_{ab}^{(\rm eff)}$ in \eqref{fe1} then describes an effective dissipative fluid with 4-velocity $u^a$. This ``fluid'' always admits (see \cite{Faraoni:2023hwu}) an imperfect fluid decomposition \cite{Faraoni:2018qdr,Pujolas:2011he} 
\be
T^{(\rm eff)} _{ab} =\rho \, u_a u_b 
+P \, h_{ab} +\pi_{ab} + 2 \, q_{(a} u_{b)} 
\,, 
\label{eq:imperfect}
\ee
where $\rho$, $P$, $\pi_{ab}$ and $q_a$ are, respectively, the effective energy density,  isotropic pressure, anisotropic stress tensor, heat flux density, $h_{ab} := g_{ab}+u_au_b$ is the induced metric onto the $3$-space orthogonal to $u^a$, and indices enclosed in parentheses are understood to be symmetrized. Explicit expressions for the fluid quantities can be obtained from the expression for $T_{ab}^{(\rm eff)}$ in \eqref{fe1} {\em via} orthogonal projections (for details see {\em e.g.} \cite{Giusti:2021sku,Miranda:2024dhw}).

Taking advantage of this representation for $T_{ab}^{(\rm eff)}$ and comparing the expressions for the fluid quantities with kinematic quantities associated with the fluid lines, one can infer some constitutive laws for the effective fluid (see \cite{Giusti:2021sku,Miranda:2024dhw}). In particular, these relations allow for an analogy with Eckart's first-order thermodynamics \cite{Eckart:1940te}, which yields a notion of temperature (multiplied by the thermal conductivity of the fluid) of modified gravity that reads
\be
\K \T = \epsilon \, \frac{\sqrt{2X}\left(G_{4\psi}-XG_{3X}\right)}{G_{4}}\label{KTdefinition} \, .
\ee
This quantity measures the departure of Hondeski gravity from GR (see \cite{Faraoni:2021lfc,Faraoni:2021jri,Giusti:2021sku,Giardino:2023ygc}, and \cite{Giusti:2021sku,Miranda:2024dhw} for the specific case of Hondeski). 

Hondeski gravity reduces to GR for $\psi = {\rm constant}$, which implies $\K \T = 0$. However, note that for GR the imperfect fluid representation of the effective fluid is not defined, hence this should be understood as a definition of the temperature of GR within the formalism. 

It is worth noting that there exist some alternatives, yet physically equivalent, definitions of such a temperature of gravity \cite{Gallerani:2024gdy} but they will not be discussed further in this work.

Combining \eqref{KTdefinition} with the field equations of viable Horndeski gravity it is also possible to derive an evolution equation for the temperature of gravity. This equation reads \cite{Miranda:2024dhw,Gallerani:2024gdy}
\begin{equation}
\begin{split}
    \frac{{\rm d}(\K \T)}{{\rm d} \tau}&=\left(\epsilon\dfrac{\Box\psi}{\sqrt{2X}}-\Theta\right)\K \T\\
    &\quad +\nabla^c\psi\nabla_c\left(\dfrac{G_{4\psi}-XG_{3X}}{G_4}\right) \, ,
\end{split}    
\label{evolution_general} 
\end{equation}
where $\tau$ stands for the proper time along the fluid lines ({\em i.e.} $\d / \d \tau := u^c \nabla_c$) and $\Theta := \nabla_c u^c$ denotes the expansion scalar. Concretely, Eq.~\eqref{evolution_general} determines whether an exact solution of a specified (viable) Horndeski model approaches GR over time. 

The ``generalised heat equation'' \eqref{evolution_general} is the basis for the thermodynamic interpretation of the attractor-to-GR mechanism for ``first-generation'' scalar-tensor theories (see \cite{Faraoni:2025alq}). Indeed, if we consider the simplified scenario of BD gravity with conformal matter ($T^{(m)} = 0$), constant BD coupling $\omega$, and a quadratic potential, then Eq.~\eqref{evolution_general} reduces to 
\be
\label{eq:FG}
\frac{\d\left( \K\T \right)}{\d\tau} = \K\T \left( \K\T -\Theta \right) \, , 
\ee
which implies that, if $\K\T > 0$ and $\Theta>0$, then: 
\begin{itemize}
\item if a solution of the BD field equations begins above the line $\K\T = \Theta$, then $\K\T \to + \infty$ --- this means that the model diverges away from GR;
\item if a solution of the BD field equations begins below the line $\K\T = \Theta$, then $\K\T \to 0$ --- thus the model converges toward the GR equilibrium state.
\end{itemize}
This discussion has been made explicit for the ``first quadrant'' of $(\Theta,\K\T)$-plane, but can easily be extended to the other quadrants. For further details we refer the reader to \cite{Faraoni:2025alq}. Note that, in contrast to earlier interpretations \cite{Faraoni:2025alq}, the line $\K \T = \Theta$ can be dynamically crossed provided that the condition $\d (\K \T) / \d \Theta = 0$ holds at the crossing point \cite{Fara20252}.
\section{The case of Induced Gravity}
\label{sec:InducedThermo}
Induced gravity \eqref{eq:CV} can easily be understood as a subclass of viable Horndeski gravity \eqref{eq:actionH} with $\psi = \phi$ and
\be 
\label{eq:H-IG}
G_4 = \xi \, \phi^2 \, , \,\,\, G_2 = 2 X - \frac{\lambda \, \phi^4}{2} \, , \,\,\, G_3 = 0 \, ,
\ee
with $2 X = - \nabla_a \phi \nabla^a \phi$. Then the temperature of (induced) gravity reads
\be 
\label{eq:temp}
\K \T = 2 \, \epsilon \, \frac{\sqrt{- \nabla_a \phi \nabla^a \phi}}{\phi} \, ,
\ee
provided that $\phi$ has a timelike gradient.

Similarly, we can compute the evolution equation for the temperature, from that of viable Horndeski \eqref{evolution_general}, which yields
\be 
\label{eq:eq-IG}
\frac{{\rm d}(\K \T)}{{\rm d} \tau}=
 \frac{\K\T}{2} (\K \T - 2 \Theta) + 2 \, \frac{\Box \phi}{\phi} \, . 
\ee
Indeed, from \eqref{eq:temp} it suffices to observe that
$$
\sqrt{2X} = \frac{\phi \, \K \T}{2 \, \epsilon} \quad \mbox{and} \quad 
\frac{(\K\T)^2}{2} = \frac{4 X}{\phi^2} \, ,
$$
hence
\begin{equation*}
\begin{split}
\frac{{\rm d}(\K \T)}{{\rm d} \tau} &= \left(\epsilon\dfrac{\Box\phi}{\sqrt{2X}}-\Theta\right)\K \T\\
    &\quad +\nabla^c\phi\nabla_c\left(\dfrac{G_{4\phi}-XG_{3X}}{G_4}\right) \\
&= \left(\epsilon\dfrac{\Box\phi}{\sqrt{2X}}-\Theta\right)\K \T +\nabla^c\phi\nabla_c\left(\frac{2}{\phi}\right) \\
&= \left(\dfrac{2 \epsilon^2 \, \Box\phi}{\phi \, \K \T}-\Theta\right)\K \T - \frac{2}{\phi^2} \, \nabla^c\phi \nabla_c\phi\\
&= \left(2 \, \dfrac{\Box\phi}{\phi \, \K \T}-\Theta\right)\K \T + \frac{4X}{\phi^2} \\
&= 2 \, \dfrac{\Box\phi}{\phi} - \Theta \, \K \T + \frac{(\K\T)^2}{2} \\
&= \frac{\K\T}{2} (\K \T - 2 \Theta) + 2 \, \frac{\Box \phi}{\phi} \, ,
\end{split}
\end{equation*}
which concludes the derivation of \eqref{eq:eq-IG}.

Due to the field equations \eqref{eq:CVeq1} we also have that
\be 
\label{eq:eq-IG-2}
\frac{{\rm d}(\K \T)}{{\rm d} \tau}=
 \frac{\K\T}{2} (\K \T - 2 \Theta) + 2 \, \lambda \, \phi^2 - 2 \, \xi \, R
 \, . 
\ee
If we require GR to be a stable equilibrium point in the far future, this implies that $\K \T \to 0$ and $\d (\K \T)/\d \tau \to 0$ must hold together as $\tau \to+\infty$ for a solution of \eqref{eq:CVeq1} and \eqref{eq:CVeq2} that relaxes to this state. In other words the following limit must hold
\be
\label{eq:necessary}
 \lim_{\tau \to + \infty} (\lambda \, \phi^2 - \xi \, R) = 0 \, ,
\ee
provided that the limit exists.

Let us now specialise the discussion to  cosmological solutions, specifically to flat FLRW spacetimes, for which the temperature associated with induced gravity  reads 
\be 
\label{eq:KT-cosmo}
\K \T = 2 \, \epsilon \, \frac{|\dot{\phi}|}{\phi} \, , 
\ee
for a scalar field $\phi$ which is not identically zero. Furthermore, using the fact that $\Theta = 3H$ and $R = 6 (\dot{H} + 2H^2)$, we can rewrite \eqref{eq:eq-IG-2} as
\begin{equation}
\label{eq:eq-IG-cosmo} 
\begin{split}
\frac{{\rm d}(\K \T)}{{\rm d} \tau} &=
 \frac{\K\T}{2} (\K \T - 6H) \\
 & \qquad + 2 \, \lambda \, \phi^2 - 12 \, \xi \, (\dot{H} + 2H^2)
 \, , 
\end{split}
\end{equation}
where in turn $H$ is related to the contribution of matter through the effective Friedmann equations for induced gravity. 

A natural observation that can be made concerns the de Sitter {\em vacuum} solution found in \cite{Cooper:1981byv} and discussed in Sec. \ref{sec:intro}. Indeed, from Eq.~\eqref{eq:dSVacuum} one immediately finds $\K \T =0$ and ${\rm d}(\K \T)/{\rm d} \tau = 0$, {\em i.e.} we are in GR and this configuration is a stable equilibrium state. This is consistent with the original analysis of \cite{Cooper:1981byv,Turchetti:1981zb} where it was noted that for such a solution the action in \eqref{eq:CV} reduces to the Einstein-Hilbert action with gravitational coupling $M_{\rm P}^2 = \xi \phi^2 _0$ plus a cosmological constant $\Lambda = \lambda \phi^4_0 /4$.

\section{More on the attractor to GR}
\label{sec:conformal}
It turns out that the effective heat equation for induced gravity \eqref{eq:eq-IG} can be further simplified if we make explicit its dependence on the stress-energy tensor of matter. 

Consider the trace of Eq.~\eqref{eq:CVeq2}, which yields
\begin{equation}
\label{eq:conto1}
- \xi \, \phi^2 \, R = T^{(\rm m)} + 2 X - \lambda \phi^4 - 3 \, \xi \, \Box \phi^2 \, . 
\end{equation}
It is easy to see that
\begin{equation}
\label{eq:conto2}
\begin{split}
\Box \phi^2 &= \nabla _c \nabla ^c \phi^2 = 
2 ( \nabla _c \phi \nabla ^c \phi + \phi \Box \phi) \\ 
&= 2 (-2X + \phi \Box \phi) \, .
\end{split}
\end{equation}
Inserting these results into Eq.~\eqref{eq:CVeq1} one finds
\begin{equation}
\label{eq:conto3}
\begin{split}
\Box \phi &= - \xi \, \phi \, R + \lambda \, \phi^3 \\
&= \frac{T^{\rm (m)}}{\phi} + \frac{2X}{\phi} - \frac{6\xi}{\phi} (-2X + \phi \Box \phi)\\
&= \frac{1}{\phi} \left[ T^{\rm (m)} + (1 + 6 \xi) \, 2X - 6\xi \, \phi \Box \phi \right] \, ,
\end{split}
\end{equation}
which implies 
\begin{equation}
(1 + 6 \xi) \left[ \phi \Box \phi - 2X\right] = T^{\rm (m)} \, ,
\end{equation}
that, taking advantage of \eqref{eq:conto2}, can be rewritten as 
\begin{equation}
\label{eq:CV1v2}
(1 + 6 \xi) \, \Box \phi^2 = 2 \, T^{\rm (m)} \, .
\end{equation}

Dividing Eq.~\eqref{eq:conto2} by $2 \,\phi^2$ one finds
\begin{equation}
\label{eq:conto4}
\frac{\Box \phi}{\phi} = \frac{2X}{\phi^2} + \frac{\Box \phi^2}{2 \, \phi^2} \, ,
\end{equation}
and recalling that $(\K \T)^2/4  = 2X/\phi^2$ and Eq.~\eqref{eq:CV1v2}, then \eqref{eq:conto4} implies
\begin{equation}
\label{eq:BoxSemplificato}
\frac{\Box \phi}{\phi} = \frac{(\K \T)^2}{4} + \frac{T^{(\rm m)}}{(1+6 \xi) \, \phi^2} \, .
\end{equation}
If we now replace $\Box \phi/\phi$ in \eqref{eq:eq-IG} with the expression in \eqref{eq:BoxSemplificato}, then the effective heat equation for induced gravity becomes
\begin{equation*}
\begin{split}
\frac{{\rm d}(\K \T)}{{\rm d} \tau} &=
\frac{\K\T}{2} (\K \T - 2 \Theta) + 2 \, \frac{\Box \phi}{\phi}\\
&= \frac{(\K\T)^2}{2} - \Theta \, \K \T 
+ 2 \, \left[ \frac{(\K \T)^2}{4} + \frac{T^{(\rm m)}}{(1+6 \xi) \, \phi^2} \right]\\
&= (\K \T)^2 - \Theta \, \K \T  + \frac{2 \, T^{(\rm m)}}{(1+6 \xi) \, \phi^2} \, ,
\end{split}
\end{equation*}
namely,
\begin{equation}
\label{eq:evolutionimproved}
\frac{{\rm d}(\K \T)}{{\rm d} \tau} = \K \T (\K \T - \Theta)  + \frac{2 \, T^{(\rm m)}}{(1+6 \xi) \, \phi^2} \, .
\end{equation}
This equation is just a rewriting of Eq.~\eqref{eq:eq-IG}, obtained by taking advantage of the field equations of induced gravity, and holds for generic solutions of \eqref{eq:CVeq1}--\eqref{eq:CVeq2} and generic matter sources. 

Notably, if we now consider conformal matter, for which $T^{\rm (m)} = 0$, the evolution equation for the temperature of induced gravity reduces to \eqref{eq:FG} and hence the existence of an attractor mechanism to GR can be determined in the same way as in \cite{Faraoni:2025alq}, and as described in Sec.~\ref{sec:thermo}. This is, after all, not surprising since induced gravity maps onto a subclass of BD gravity analysed in detail in \cite{Faraoni:2025alq} ($\omega = {\rm const.}$ and $\psi V' - 2V = 0$).

From a cosmological perspective, if we now consider an expanding universe with ordinary matter (dust or radiation), then $T^{\rm (m)} \to 0$ as $t \to + \infty$ since $T^{\rm (m)}$ is either identically zero (radiation) or matter dilutes as the universe expands (dust, $T^{\rm (m)} = - \rho^{\rm (m)} \propto -a^{-3}$). This is consistent with the necessary condition \eqref{eq:necessary} discussed in Sec.~\ref{sec:InducedThermo} for the existence of the GR attractor in the far future.

\section{Conclusions}
\label{sec:conc}
In this work we have revisited the induced gravity model based on the thermodynamics of scalar-tensor gravity. Specifically, we have taken advantage of the fact that the action proposed by Cooper and Venturi \cite{Cooper:1981byv} belongs to a subclass of viable Horndeski theories, for which the thermodynamic formalism was developed in general in \cite{Giusti:2021sku}, and specialised to cosmology in \cite{Miranda:2024dhw}. In particular, we found an explicit expression for both the temperature of gravity \eqref{eq:temp} and its evolution equation \eqref{eq:evolutionimproved}. These results allowed us to reinterpret the scale-invariance-breaking de Sitter {\em vacuum} solution of induced gravity as a stable zero-temperature equilibrium state of the effective dissipative fluid associated with the scalar field $\phi$. Furthermore, taking advantage of the analysis in \cite{Faraoni:2025alq}, we identified a necessary condition [Eq.~\eqref{eq:necessary}] for the existence of a stable GR equilibrium state at late times, {\em i.e.} an attractor mechanism to GR (with cosmological constant). Notably, in the case of conformal matter the analysis of the attractor-to-GR mechanism reduces exactly to that of \cite{Faraoni:2025alq}, therefore leading to the same conclusions (summarised in Sec. \ref{sec:thermo}).

Finding exact solutions upon which one could test this general analysis is rather difficult due to the complexity of the field equations. However, some solutions (both {\em vacuum} and with matter) have been investigated numerically \cite{Finelli:2007wb} or for anisotropic cosmologies \cite{Kamenshchik:2017ojc}. These solutions will be the focus of future investigations.
\section*{Acknowledgments}
\noindent The author is grateful to A. Tronconi for helpful comments that greatly improved the manuscript, and to R. Casadio, L. Gallerani, L. Caraffi, M. Miranda, A. Mentrelli, and V. Faraoni for valuable discussions.

\noindent The author is supported by the Italian Ministry of Universities and Research (MUR) through the grant ``BACHQ: Black Holes and The Quantum'' (grant no. J33C24003220006) and by the INFN grant FLAG.

\noindent This work has been carried out in the framework of activities of the National Group of Mathematical Physics (GNFM, INdAM).
%
%
%%% Bibliography
%
%\bibliographystyle{apsrev4-2}
%\bibliography{Biblio}

%
\end{document}